\documentclass[twocolumn,fullpaper]{jpsj3}

\usepackage{graphicx}% Include figure files
\usepackage{dcolumn}% Align table columns on decimal point
\usepackage{bm}% bold math
\usepackage{color}
\usepackage{ulem}
\usepackage{booktabs}

\begin{document}

\title{Antiferromagnetic Order and Magnetic Frustration in the Honeycomb Heavy-Fermion System Ce(Pt$_{1-x}$Pd$_{x}$)$_6$Al$_3$: $^{27}$Al and $^{195}$Pt NMR Studies}

\author{Shunsaku~Kitagawa$^{1,}$\thanks{E-mail address: kitagawa.shunsaku.8u@kyoto-u.ac.jp}, 
Fumiya~Hori$^{1,}$\thanks{Present address: Department of Physics, Tohoku University, Sendai 980-8578, Japan}, 
Kenji~Ishida$^{1}$,
Ryohei~Oishi$^{2,}$\thanks{Present address: Research Institute for Electronic Science, Hokkaido University, Sapporo 001-0020, Japan},
Yasuyuki~Shimura$^{2}$, 
Takahiro~Onimaru$^{2}$, and
Toshiro~Takabatake$^{2}$
}

\inst{$^1$Department of Physics, Graduate School of Science, Kyoto University, Kyoto 606-8502, Japan \\
$^3$Department of Quantum Matter, Graduate School of Advanced Science and Engineering, Hiroshima University, Higashi-Hiroshima 739-8530, Japan
}

\date{\today}

\abst{
Heavy-fermion systems with magnetic frustration offer a rich platform for investigating the interplay among Kondo screening, magnetic frustration, and quantum criticality.
We report comprehensive $^{27}$Al and $^{195}$Pt nuclear magnetic resonance measurements on polycrystalline Ce(Pt$_{1-x}$Pd$_{x}$)$_6$Al$_3$ ($x = 0$, 0.1, 0.2, and 0.3).
For $x = 0$, the Knight shift, linewidth, and nuclear spin–lattice relaxation rate reveal a paramagnetic heavy-fermion ground state persisting down to 0.1~K, characterized by a coherence temperature $T_{\mathrm{coh}} \simeq 15$~K.
Substituting Pd induces antiferromagnetic order at $T_{\mathrm{N}} \simeq 3.5$~K, while suppressing $T_{\mathrm{coh}}$.
Comparison between $x = 0.1$ and $x = 0.3$ reveals a crossover from itinerant spin-density-wave antiferromagnetism to more localized-moment antiferromagnetism, indicating a shift toward the localized side of the Doniach phase diagram.
These findings establish Ce(Pt$_{1-x}$Pd$_{x}$)$_6$Al$_3$ as a tunable platform to explore the competition between Kondo screening and magnetic frustration.
}

\maketitle

\section{Introduction}

Quantum materials in which several competing interactions coexist offer rich grounds for exploring nontrivial emergent phenomena such as quantum criticality and quantum spin-liquid (QSL) states~\cite{Anderson1973,Balents2010,Zhou2017Quantum}.
One canonical route to such competition arises when nearest-neighbor ($J_{1}$) and next-nearest-neighbor ($J_{2}$) exchange interactions are comparable in magnitude, leading to magnetic frustration.
In insulating magnets, the realization of QSL states due to magnetic frustration has been actively studied in triangular, kagom\'e, zigzag or honeycomb networks~\cite{Obradors1988,PhysRevLett.91.107001,Nakatsuji2005,PhysRevLett.99.137207,Smirnova_JACS_2009,PhysRevX.1.021002,Yan2011,Han2012,S.Kitagawa_JPSJ_2015,Fu2015Evidence,Kawasugi2023,Iwase2023,Eto_JPSJ_2023,Kogre_JPSJ_2023,Nihongi_JPSJ_2024,Hori2024}.
Recent theoretical and experimental studies have demonstrated that considering anisotropic magnetic interactions can lead to the emergence of novel low-energy excitations, such as Majorana fermions and nematic quasiparticles~\cite{Kitaev2006,Shen2016,Banerjee2017,Paddison2017Continuous,Kasahara2018Majorana,Kitagawa2018SpinOrbital,Hori2023,Yoshimoto_JPSJ_2023,Yamada_JPSJ_2023,Saito_PRL_2024,Sato_JPSJ_2024,Saito_JPSJ_2024}.

While many investigations focus on localized spins without conduction electrons, the interplay between localized $f$-electrons and conduction carriers in metallic systems represents an equally fascinating frontier~\cite{Yang_PRB_2017,Kuchler_PRB_2017}.
In $f$-electron compounds, theoretical studies have predicted that tuning the ratio of the Kondo coupling $J_{cf}$ to the intersite magnetic interactions $J_{1}$ and $J_{2}$ can yield a wide variety of ground states, ranging from heavy Fermi liquids to magnetically ordered phases and possibly metallic QSLs~\cite{Coleman2010,PhysRevB.107.205151,Tokiwa2014}.
Moreover, it is well established that many heavy-fermion compounds exhibit unconventional superconductivity in the vicinity of a magnetic quantum critical point (QCP)~\cite{F.Steglich_PRL_1979,H.R.Ott_PRL_1983,G.R.Stewart_PRL_1984,C.Pfleiderer_RMP_2009,D.Aoki_JPCondMatt_2022,S.Kitagawa_JPSJ_2024}.
In recent years, increasing attention has been drawn to superconductors on frustrated systems, such as kagom\'e networks, in which unconventional pairing and ordered states have been reported~\cite{Jrome1991,Takada2003,S.Kitagawa_JPSJ_2020_b,PhysRevLett.125.247002,wangElectronicNatureChiral2021,xiangTwofoldSymmetry$c$axis2021,fukushimaViolationEmergentRotational2024}.
From this perspective, systems that host both strong Kondo interactions and magnetic frustration offer a unique platform to explore novel quantum phases.

\begin{figure}[!tb]
\centering
\includegraphics[width=8.5cm,clip]{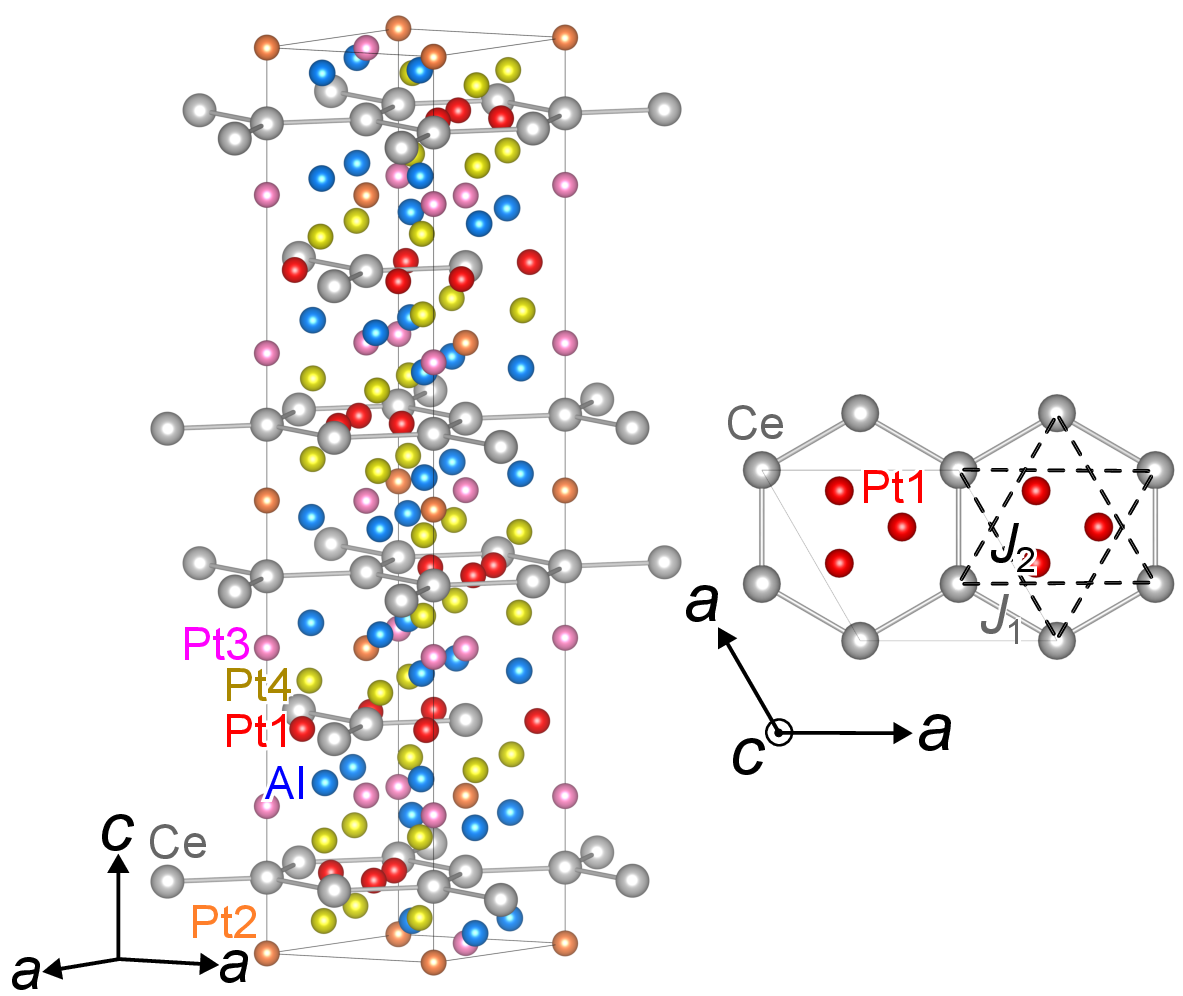}
\caption{
(Color online) Crystal structure of CePt$_6$Al$_3$ drawn by VESTA~\cite{K.Momma_JAC_2011}.
A box indicates the unit cell.
Ce atoms form a two-dimensional honeycomb network.
There are four inequivalent Pt sites, whereas Al occupies a single crystallographic site.
We represent nearest-neighbor interaction $J_{1}$ and next-nearest-neighbor interaction $J_{2}$ in the right panel.
}
\label{Fig.1}
\end{figure}

\begin{figure*}[!tb]
\centering
\includegraphics[width=\linewidth,clip]{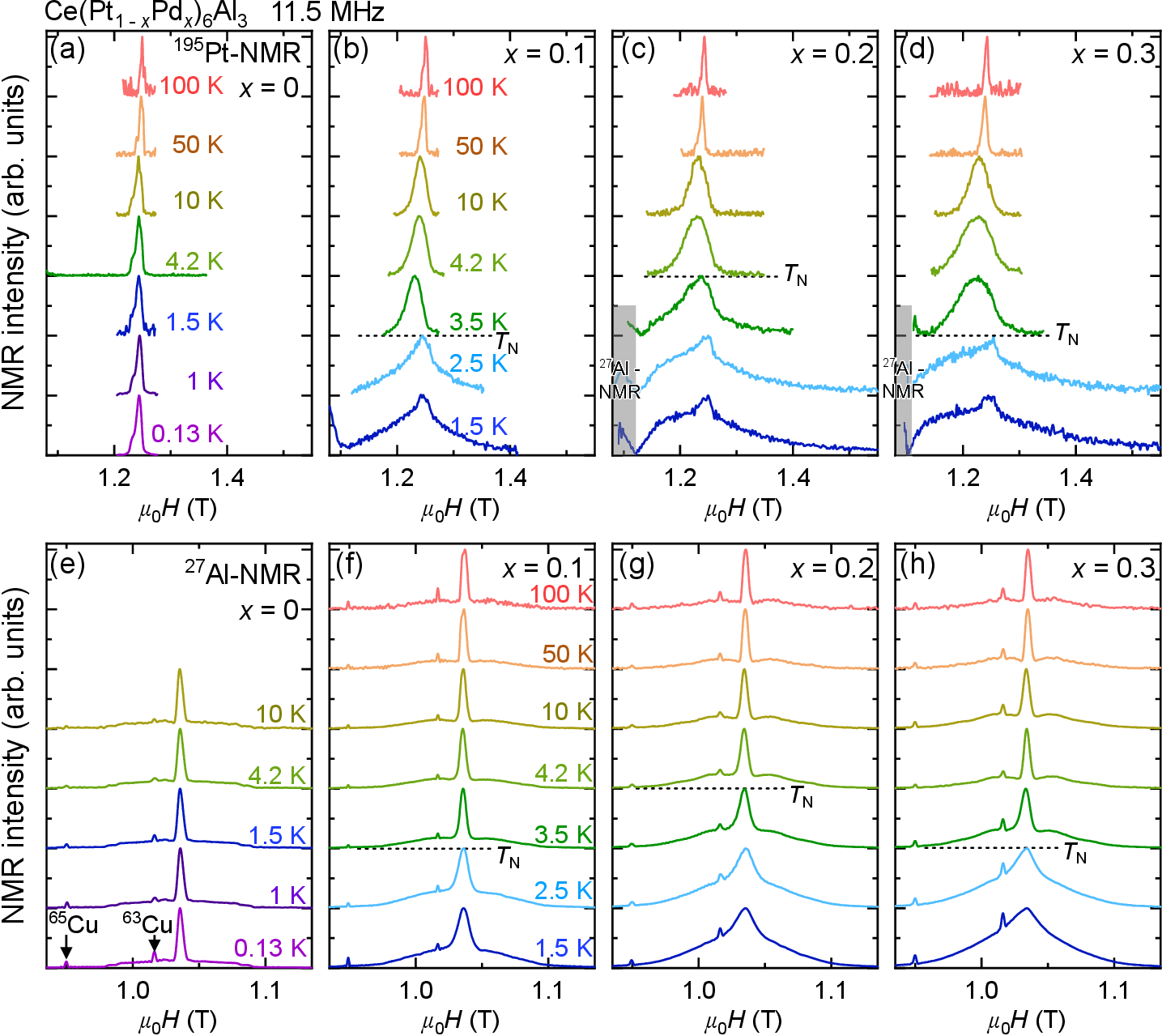}
\caption{
(Color online) Temperature variation of the NMR spectra for Ce(Pt$_{1-x}$Pd$_{x}$)$_6$Al$_3$ measured at 11.5~MHz.
(a)--(d): $^{195}$Pt NMR spectra for $x=0$, 0.1, 0.2, and 0.3, respectively.
(e)--(h): $^{27}$Al NMR spectra for $x=0$, 0.1, 0.2, and 0.3, respectively.
}
\label{Fig.2}
\end{figure*}

Ce(Pt$_{1-x}$Pd$_{x}$)$_6$Al$_3$ is one of these candidate systems.
CePt$_6$Al$_3$ crystallizes in the NdPt$_6$Al$_3$-type trigonal structure with a centrosymmetric space group $R\bar{3}c$ ($D_{3d}^{6}$, No.~167), in which Ce atoms form a two-dimensional honeycomb network (Fig.~\ref{Fig.1})~\cite{Eustermann2017}.
This honeycomb arrangement naturally leads to magnetic frustration among the localized $4f$ moments due to the competing interactions between $J_1$ and $J_2$.
In related compounds where Ce is replaced by other rare-earth elements, a variety of magnetic structures have been observed~\cite{Oishi_JPSJ_2022,Oishi_PRB_2024,Oishi_JPSJ_2024}.
The measurements of specific heat, magnetic susceptibility, and electrical resistivity on CePt$_6$Al$_3$ indicate the formation of a heavy-electron state below Kondo temperature $T_{\mathrm{K}} \sim 10$~K, suggesting a strong competition between Kondo screening and frustrated magnetic interactions~\cite{Oishi_JPSJ_2020}.

\begin{figure*}[!tb]
\centering
\includegraphics[width=14cm,clip]{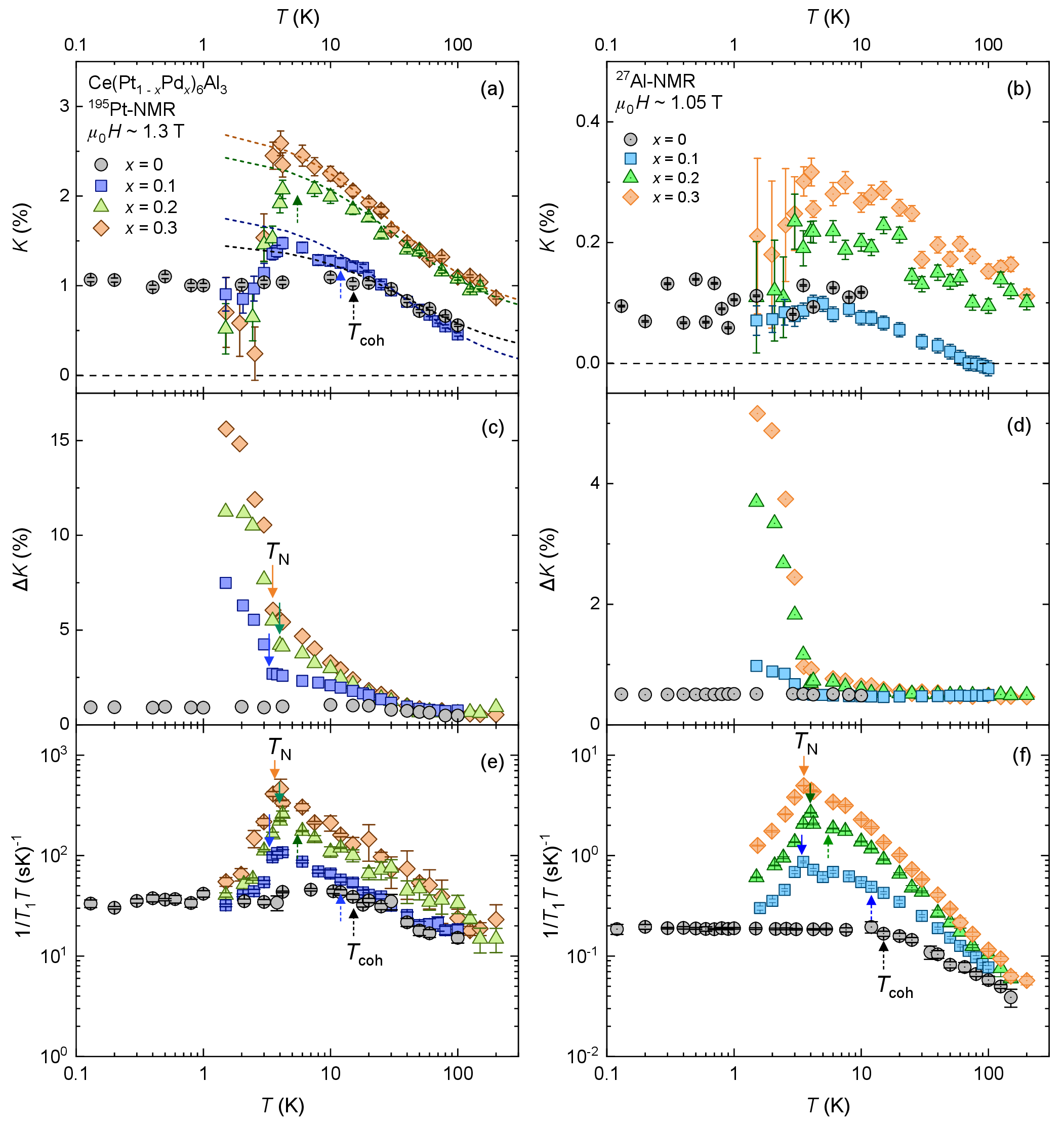}
\caption{
(Color online) Temperature dependence of the NMR quantities for Ce(Pt$_{1-x}$Pd$_{x}$)$_6$Al$_3$.
(a), (b): Knight shift $K$ for (a) $^{195}$Pt and (b) $^{27}$Al.
The broken curves indicate Curie-Weiss behavior.
(c), (d): Linewidth $\Delta K$ for (c) $^{195}$Pt and (d) $^{27}$Al.
(e), (f): Nuclear spin-lattice relaxation rate $1/T_1T$ for (e) $^{195}$Pt and (f) $^{27}$Al.
The suppression of $1/T_1T$ and the saturation of $K$ at low temperatures signal the formation of heavy-fermion coherence, whereas the anomalies at $T_{\mathrm{N}}$ indicate the onset of antiferromagnetic order.
The solid (dashed) arrows indicate $T_{\rm N}$ $(T_{\rm coh})$.
}
\label{Fig.3}
\end{figure*}

Partial substitution of Pt by Pd in CePt$_6$Al$_3$ retains the trivalent state of Ce, while systematically reducing the absolute value of the paramagnetic Curie temperature $\theta_{\mathrm{p}}$~\cite{Oishi_PRB_2021,Oishi_JPCS_2022}.
At substitution levels $x \ge 0.1$, a long-range antiferromagnetic (AFM) order emerges, implying that Pd doping reduces Kondo coupling $J_{cf}$ or/and relieves magnetic frustration.
Specific heat measurements reveal that the entropy recovered just above the N\'eel temperature $T_{\mathrm{N}}$ increases with $x$ and that the estimated $T_{\mathrm{K}}$ decreases from $\sim 10$~K at $x = 0$ to $\sim 3.5$~K at $x = 0.3$.
These results suggest a gradual suppression of the Kondo effect with Pd substitution~\cite{Oishi_PRB_2021}.

Furthermore, the frustration parameter $f = |\theta_{\mathrm{p}}|/T_{\mathrm{N}}$ decreases with increasing $x$, although care must be taken because both $\theta_{\mathrm{p}}$ and $T_{\mathrm{N}}$ can be influenced by the Kondo effect.
The sign of resistivity change below $T_{\mathrm{N}}$ indicates a difference in the nature of the antiferromagnetism: At $x = 0.1$, a spin-density-wave (SDW)–like AFM order is realized, whereas at $x \ge 0.2$, the behavior is more consistent with a localized-moment type of AFM order~\cite{Oishi_PRB_2021}.

In this work, to microscopically investigate how the magnetic properties evolve with Pd substitution, we performed $^{27}$Al and $^{195}$Pt nuclear magnetic resonance (NMR) measurements on polycrystalline Ce(Pt$_{1-x}$Pd$_{x}$)$_6$Al$_3$ for $x = 0$, 0.1, 0.2, and 0.3.
Our results indicate that CePt$_6$Al$_3$ remains paramagnetic down to $0.1$~K, while Pd substitution induces AFM order.
By comparing the results at $x = 0.1$ with those at $x = 0.3$, we identify a crossover from itinerant to localized magnetism as the Kondo coupling is suppressed, which is consistent with the bulk measurements~\cite{Oishi_PRB_2021,Oishi_JPCS_2022}.
These findings demonstrate that the competition between Kondo screening and magnetic frustration governs the evolution of the ground state in Ce(Pt$_{1-x}$Pd$_{x}$)$_6$Al$_3$, providing a unique platform to explore quantum criticality.

\section{Experimental}

Polycrystalline samples of Ce(Pt$_{1-x}$Pd$_{x}$)$_6$Al$_3$ with $x = 0$, 0.1, 0.2, and 0.3 were synthesized by a conventional arc-melting method followed by annealing~\cite{Oishi_PRB_2021}.
The Pd composition $x$ determined by wavelength-dispersive electron-probe microanalysis for the main phase did not deviate from the nominal value $x$ within the resolution.
$^{27}$Al (nuclear spin $I = 5/2$, gyromagnetic ratio $^{27}\gamma/2\pi = 11.094$~MHz/T, nuclear quadrupole moment $^{27}Q = 0.1466 \times 10^{-28}$~m$^2$) and $^{195}$Pt ($I = 1/2$, $^{195}\gamma/2\pi = 9.153$~MHz/T) NMR measurements were carried out using a conventional spin-echo technique~\cite{R.K.Harris_2001,N.J.Stone_Q_2016}.
The magnetic field-sweep NMR spectra were obtained by recording the spin-echo signal observed after a standard $\pi/2$--$\pi$ radio frequency pulse sequence at 11.5~MHz.
The magnetic field was calibrated using a $^{63/65}$Cu [$^{63(65)}\gamma/2\pi = 11.285 (12.089)$~MHz/T]-NMR signal with the Knight shift $K_{\rm Cu} = 0.2385$\% from a NMR coil~\cite{Metallicshifts_1977}.
Knight shift was determined from the peak position of the NMR spectrum, and the full width at half-maximum (FWHM), $\Delta K$, was used to characterize the spectral linewidth.
The nuclear spin-lattice relaxation rate $1/T_1$ was measured at the central peak of the NMR spectra.
$1/T_1$ was evaluated by fitting the relaxation curve of the nuclear magnetization after saturation to a theoretical function for the nuclear spin $I = 1/2$ ($^{195}$Pt), which is a single exponential function, and that for $I = 5/2$ ($^{27}$Al).
Temperature control down to 0.1~K was achieved using a $^3$He–$^4$He dilution refrigerator.

\section{Results and Discussion}

\subsection{Static magnetic properties}
Figure~\ref{Fig.2} presents the $^{195}$Pt- and $^{27}$Al-NMR spectra for Ce(Pt$_{1-x}$Pd$_{x}$)$_6$Al$_3$ with $x = 0$, 0.1, 0.2, and 0.3 over a wide temperature range.
The $^{27}$Al NMR spectra are relatively sharp and exhibit well-resolved satellite peaks, reflecting the existence of electric field gradients.
In contrast, the $^{195}$Pt NMR spectra are considerably broader and complex shapes even in the paramagnetic state, which can be attributed to the presence of four crystallographically inequivalent Pt sites, as shown in Fig.~\ref{Fig.1}.
The line widths of the $^{27}$Al and $^{195}$Pt spectra are almost unchanged with increasing Pd concentration.
Upon cooling, the linewidth of $^{195}$Pt grows more markedly than that of $^{27}$Al, indicating a larger hyperfine coupling constant for Pt and the presence of significant magnetic anisotropy.
The difference in the hyperfine coupling constants may originate from the distances from the Ce atoms, as shown in Fig.~\ref{Fig.1}.

For $x \ge 0.1$, a pronounced broadening of the spectra is observed below approximately 3.5~K, signaling the onset of long-range AFM order.
In contrast, no discernible broadening is detected for $x = 0$ down to 0.1~K, indicating the absence of magnetic order.
In the ordered state, the $^{27}$Al NMR spectra broaden symmetrically, whereas the $^{195}$Pt spectra exhibit an asymmetric broadening.
This asymmetry likely originates from the different magnitudes and directions of internal magnetic fields at the four distinct Pt sites.
However, because the present measurements were performed on powdered samples, detailed information about the magnetic structure could not be determined.

The temperature dependence of the Knight shift $K$, determined from the peak position of the NMR spectrum, is summarized in Figs.~\ref{Fig.3}(a) and \ref{Fig.3}(b).
At high temperatures, $^{195}K$ for all compositions follows a Curie–Weiss behavior, which reflects the dominance of localized moment contributions.
On cooling, $^{195}K$ levels off below the coherence temperature $T_{\mathrm{coh}}$, indicating the formation of a coherent heavy-fermion state.
Here, $T_{\mathrm{coh}}$ is defined as the characteristic temperature below which the $^{195}K$ Knight shift begins to deviate from Curie–Weiss behavior.
For $x = 0$, $T_{\mathrm{coh}}$ is estimated to be approximately 15~K, whereas it decreases to $\sim$ 12.5~K for $x = 0.1$.
Although no distinct anomaly is observed at $x = 0.2$, there is a subtle deviation near $T_{\mathrm{K}}$ estimated from the magnetic entropy ($\sim 4.5$~K).
The suppression of the Kondo effect with Pd substitution is consistent with the previous measurements~\cite{Oishi_PRB_2021}.
Note that, as discussed later, an anomaly at around $T_{\mathrm{coh}}$ is also observed in $1/T_1T$, particularly for $^{27}$Al, although the anomaly in $^{195}K$ at $T_{\mathrm{coh}}$ is not pronounced.

\begin{figure}[!tb]
\centering
\includegraphics[width=6.5cm,clip]{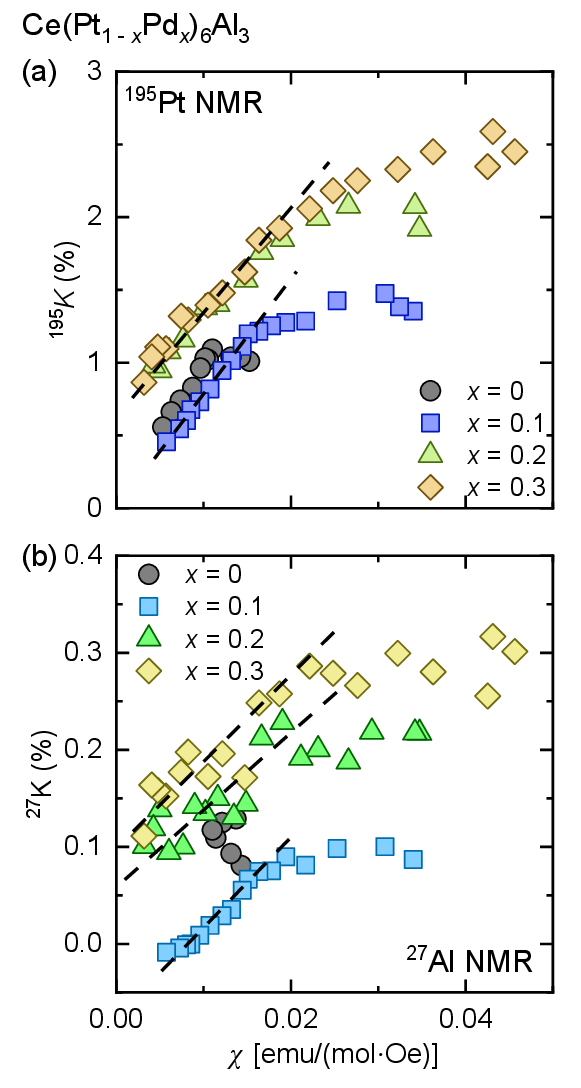}
\caption{
(Color online) $K$--$\chi$ plot for (a) $^{195}$Pt- and (b) $^{27}$Al-NMR in Ce(Pt$_{1-x}$Pd$_{x}$)$_6$Al$_3$.
The magnetic susceptibility data were obtained from Ref. 56.
The broken lines indicate the guide for the eyes.
}
\label{Fig.3-2}
\end{figure}

The Knight shift is proportional to the bulk magnetic susceptibility $\chi$ through the hyperfine interaction, as expressed by the relation,
\begin{align}
K = \frac{A_{\rm hf}}{N_{\rm A}\mu_{\rm B}} \chi + K_{0},
\label{eq.K}
\end{align}
where $A_{\mathrm{hf}}$ is the hyperfine coupling constant, $N_{\rm A}$ is the Avogadro constant, $\mu_{\rm B}$ is the Bohr magneton, and $K_0$ is the temperature-independent component of the Knight shift.
The value of $A_{\mathrm{hf}}$ can be determined from the slope of the $K$–$\chi$ plot shown in Figs.~\ref{Fig.3-2}(a) and ~\ref{Fig.3-2}(b).
For all values of $x$, a change in slope is observed around $\chi \sim 0.02$~[emu/(mol$\cdot$Oe)].
This is because, at low temperatures, the magnetic susceptibility is affected by magnetic impurities~\cite{Oishi_PRB_2021}, whereas the Knight shift selectively probes the intrinsic electronic properties in the sample.
Therefore, $A_{\mathrm{hf}}$ was determined by fitting a linear relation in the high-temperature region, where both $K$ and $\chi$ are small.
The estimated $A_{\mathrm{hf}}$ and $K_0$ are summarized in Table~I.
Note that, as the present measurements were performed on polycrystalline samples, the estimated $A_{\mathrm{hf}}$ reflects the average values.
As shown in Table~I, $A_{\mathrm{hf}}$ at the Pt site is approximately 9 times larger than that at the Al site.

\begin{table}[htbp]
\centering
\caption{
Hyperfine coupling constant $A_{\mathrm{hf}}$ (in T/$\mu_{\mathrm{B}}$) and temperature-independent Knight shift $K_0$ (in \%) in Ce(Pt$_{1-x}$Pd$_{x}$)$_6$Al$_3$}
\begin{tabular}{c c c c c}
\toprule
$x$ & Nucleus & $T$ (K) & $A_{\mathrm{hf}}$ (T/$\mu_{\mathrm{B}}$) & $K_0$ (\%) \\
\midrule
0.0 & $^{195}$Pt & $\ge 10$ & 0.52 $\pm$ 0.04 & 0.04 $\pm$ 0.06 \\
\addlinespace
0.1 & $^{195}$Pt & $\ge 18$ & 0.44 $\pm$ 0.01 & -0.02 $\pm$ 0.05 \\
    & $^{27}$Al  & $\ge 18$ & 0.044 $\pm$ 0.01  & -0.064 $\pm$ 0.007 \\
\addlinespace
0.2 & $^{195}$Pt & $\ge 15$ & 0.34 $\pm$ 0.01 & 0.69 $\pm$ 0.03 \\
    & $^{27}$Al  & $\ge 15$ & 0.038 $\pm$ 0.008 & 0.07 $\pm$ 0.02 \\
\addlinespace
0.3 & $^{195}$Pt & $\ge 15$ & 0.34 $\pm$ 0.02 & 0.78 $\pm$ 0.03 \\
    & $^{27}$Al  & $\ge 15$ & 0.041 $\pm$ 0.006 & 0.11 $\pm$ 0.01 \\
\bottomrule
\end{tabular}
\end{table}

Figures~\ref{Fig.3}(c) and \ref{Fig.3}(d) show the temperature dependence of FWHM of the NMR spectrum, $\Delta K$.
For $x \ge 0.1$, a sharp increase in $\Delta K$ is observed below $T_{\mathrm{N}} \simeq 3.5$~K, consistent with the development of static staggered internal fields associated with AFM order.
In contrast, for $x = 0$, no such anomaly in $\Delta K$ is detected down to 0.1~K, reinforcing the absence of magnetic order.
As mentioned above, the larger sensitivity of the $^{195}$Pt linewidth compared to that of $^{27}$Al can be attributed to the larger hyperfine coupling constants at the Pt sites.
These observations collectively confirm that Pd substitution induces AFM order and simultaneously suppresses $T_{\rm coh}$ in Ce(Pt$_{1-x}$Pd$_{x}$)$_6$Al$_3$.

\subsection{Dynamic magnetic properties}
Figures~\ref{Fig.3}(e) and \ref{Fig.3}(f) show the temperature dependence of $1/T_1T$ for Ce(Pt$_{1-x}$Pd$_{x}$)$_6$Al$_3$ with $x = 0$, 0.1, 0.2, and 0.3.
$1/T_1T$ exhibits a systematic variation with $x$ and appears to show little dependence on the applied magnetic field, in contrast to the NMR spectrum.
At temperatures above 100~K, $1/T_1T$ of $^{195}$Pt is nearly temperature independent for all samples, indicating a dominant contribution from conduction electrons rather than localized $4f$ moments.
With decreasing temperature, $1/T_1T$ increases approximately following a $1/T$ behavior, consistent with the development of magnetic fluctuations associated with localized moments.
Below $T_{\mathrm{coh}}$, $1/T_1T$ either saturates (for $x = 0$) or shows a change in slope (for $x = 0.1$ and 0.2), reflecting the crossover to a heavy-fermion state~\cite{Kawasaki2020}, similar to the Knight shift.

For samples with $x \ge 0.1$, a divergence of $1/T_1T$ is observed upon approaching $T_{\mathrm{N}} \simeq 3.5$~K, signaling critical slowing down of magnetic fluctuations.
Below $T_{\mathrm{N}}$, $1/T_1T$ drops sharply, indicating the opening of a gap in the magnetic excitation spectrum due to the establishment of AFM order.
The magnitude of $1/T_1T$ at the $^{195}$Pt sites is about two orders of magnitude larger than that at the $^{27}$Al sites, which can also be attributed to the stronger hyperfine coupling constant at the Pt sites.
These results are fully consistent with the Knight shift and linewidth analyses and further support the scenario that Pd substitution induces a transition from a paramagnetic heavy-fermion state to an antiferromagnetically ordered state in Ce(Pt$_{1-x}$Pd$_{x}$)$_6$Al$_3$.

\begin{figure}[!tb]
\centering
\includegraphics[width=\linewidth,clip]{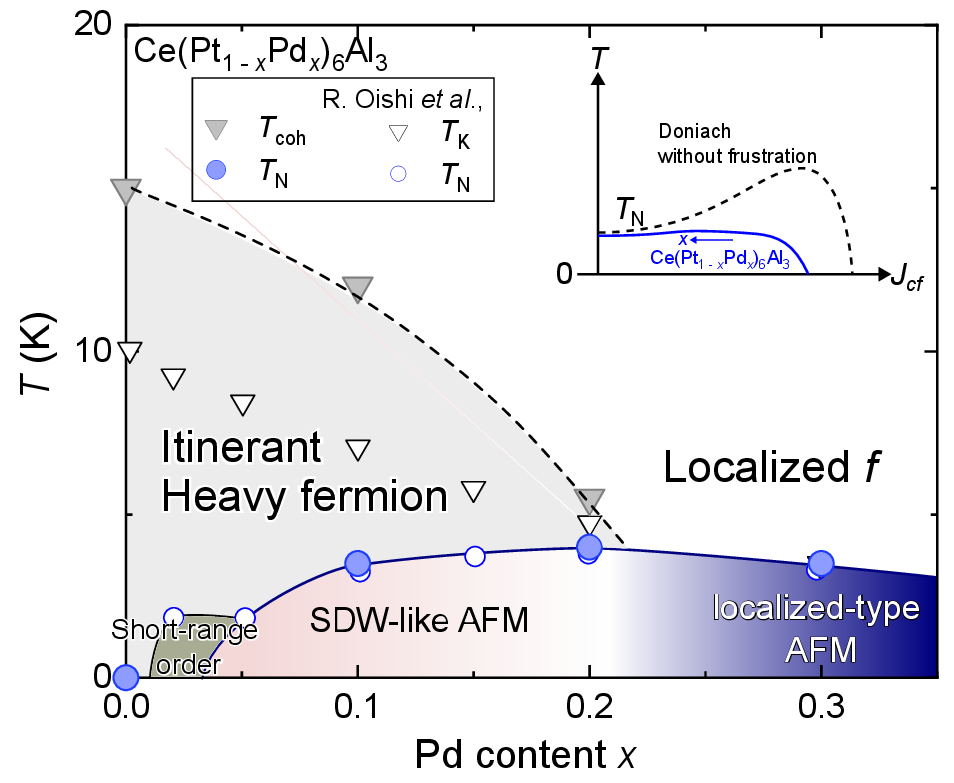}
\caption{
(Color online) Pd concentration dependence of the N\'eel temperature $T_{\mathrm{N}}$ and the coherence temperature $T_{\mathrm{coh}}$ determined from the NMR measurements.
$T_{\mathrm{N}}$ and Kondo temperature $T_{\mathrm{K}}$ determined from the magnetic susceptibility and specific heat measurements are also shown~\cite{Oishi_PRB_2021}.
(Inset) Schematic image of the magnetic phase diagram of Ce(Pt$_{1-x}$Pd$_{x}$)$_6$Al$_3$ compared with the conventional Doniach phase diagram.
}
\label{Fig.4}
\end{figure}

\subsection{Compositional Phase Diagram}
Figure~\ref{Fig.4} summarizes the Pd concentration dependence of $T_{\mathrm{coh}}$ and $T_{\mathrm{N}}$ determined from the NMR measurements.
$T_{\mathrm{coh}}$ decreases monotonically from approximately 15~K at $x = 0$ and becomes undetectable at $x = 0.3$, reflecting a systematic suppression of the Kondo coupling $J_{cf}$ with increasing Pd content.
In contrast, $T_{\mathrm{N}}$ remains nearly constant at around 3.5~K for $0.1 \leq x \leq 0.3$, while it abruptly vanishes with decreasing $x$ from 0.1 to 0, suggesting the existence of a QCP near $x_{\mathrm{c}} \simeq 0$.
Short-range magnetic order was reported at $x = 0.025$ in magnetic susceptibility measurements~\cite{Oishi_PRB_2021}.

The contrasting trends between $T_{\mathrm{coh}}$ and $T_{\mathrm{N}}$ suggest that, in addition to the reduction of $J_{cf}$, other factors such as magnetic frustration may play a significant role in stabilizing the AFM order. 
Within the conventional Doniach framework~\cite{S.Doniach_PhysicaBC_1977}, decreasing $J_{cf}$ is expected to first enhance and then suppress $T_{\mathrm{N}}$ as the system moves away from the QCP.
As a results, $T_{\mathrm{N}}$ of prototypical heavy-fermion systems such as CeCu$_{6-x}$Au$_x$~\cite{Lhneysen1996} and CeNiGe$_3$ under pressure~\cite{S.Kitagawa_JPSJ_2020} exhibits a clear dome-shaped dependence on control parameters.
In contrast, $T_{\mathrm{N}}$ in Ce(Pt$_{1-x}$Pd$_{x}$)$_6$Al$_3$ remains nearly constant for $x \geq 0.1$.
One possible scenario that explains this unconventional change in $T_{\mathrm{N}}(x)$ is that in $x = 0$, the system lies in a region where long-range AFM order would be expected without frustration.
However, strong competing interactions among Ce moments suppress the ordering temperature to zero, effectively stabilizing a QCP. 
As Pd is substituted, $J_{cf}$ decreases, which would typically reduce $T_{\mathrm{N}}$; at the same time, the Pd substitution may relieve magnetic frustration. 
In this regime, the opposing effects of reduced $J_{cf}$ (suppressing $T_{\mathrm{N}}$) and weakened frustration (enhancing $T_{\mathrm{N}}$) may effectively compensate for each other, leading to an almost constant $T_{\mathrm{N}}$ for $x \geq 0.1$, as shown in the inset of Fig.~\ref{Fig.4}.
This scenario is consistent with the systematic decrease in the frustration parameter $f = |\theta_{\mathrm{p}}|/T_{\mathrm{N}}$ as a function of $x$ observed in the previous measurements~\cite{Oishi_PRB_2021}.
The weakening of magnetic frustration with increasing Pd content may originate from a reduction in the competition between $J_1$ and $J_2$ interactions.
In addition, an enhancement of interlayer coupling, suggested by the decrease in the $c$-axis length with Pd substitution~\cite{Oishi_PRB_2021}, reduces the two-dimensionality of the electronic state, which would also act to suppress magnetic frustration.

\begin{figure}[!tb]
\centering
\includegraphics[width=\linewidth,clip]{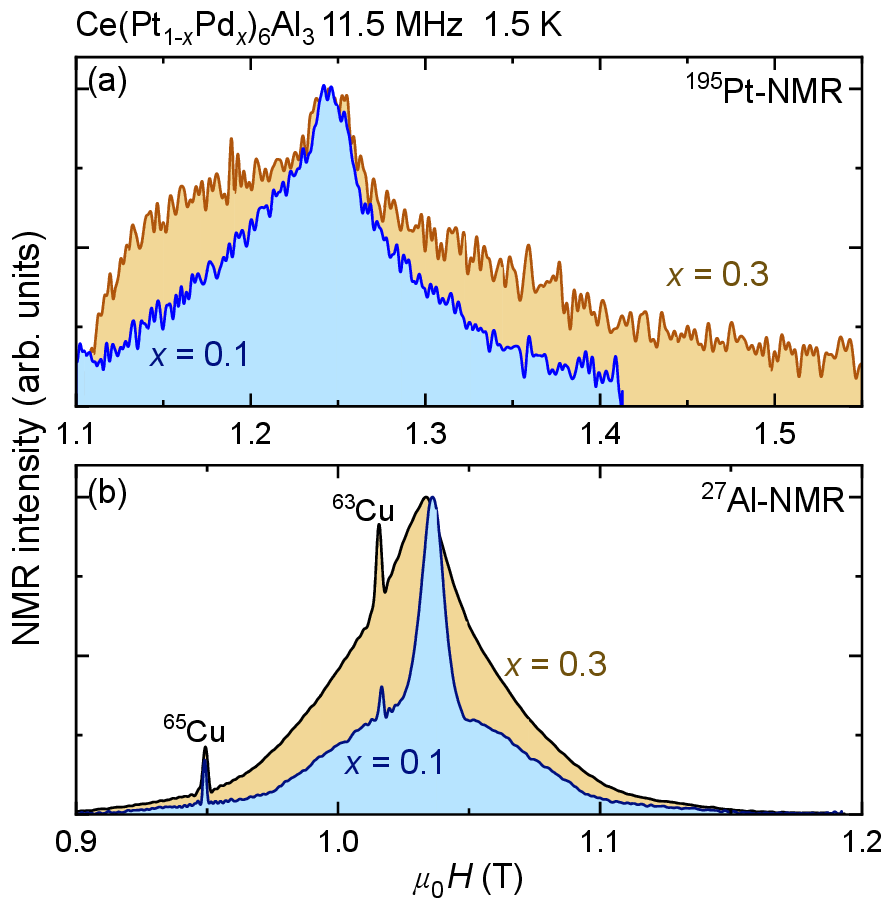}
\caption{
(Color online) Comparison of (a) $^{195}$Pt-NMR and (b) $^{27}$Al-NMR spectra at 1.5 K for $x = 0.1$ and $x = 0.3$ of Ce(Pt$_{1-x}$Pd$_{x}$)$_6$Al$_3$.
}
\label{Fig.6}
\end{figure}

\subsection{Nature of the Antiferromagnetic Order}
Comparison between the results at $x = 0.1$ and those at $x = 0.2$ and $0.3$ reveals a notable change in the character of the AFM order.
For $x = 0.1$, the presence of a residual heavy-fermion coherence just above $T_{\mathrm{N}}$, as indicated by the saturation of the Knight shift and $1/T_1T$, suggests an itinerant SDW–type AFM order.
In this regime, the ordered moments are expected to be relatively small, and Fermi surface instabilities predominantly drive the magnetism.
In contrast, for $x = 0.3$, no clear signature of heavy-fermion coherence is observed above $T_{\mathrm{N}}$.
The Knight shift and $1/T_1T$ values are larger in the paramagnetic state compared to $x = 0.1$, indicating enhanced fluctuating local moments.

In addition, as shown in Fig.~\ref{Fig.6}, both the $^{27}$Al and $^{195}$Pt spectra for $x = 0.3$ broaden significantly compared to those for $x = 0.1$, suggesting an increase in the ordered moment.
While structural features in the $^{195}$Pt spectra are also observed, because of the powder-sample measurements and the presence of multiple crystallographically distinct Pt sites, it is difficult to make definitive conclusions regarding changes in magnetic structure.
These observations suggest that the antiferromagnetism in the $x = 0.3$ sample is more localized, with larger ordered moment associated with the $4f$ electrons.
Thus, Pd substitution drives the system from an itinerant SDW regime at $x = 0.1$ toward a localized moment AFM regime for $x = 0.3$.
This evolution is consistent with a movement toward the localized side of the Doniach phase diagram as the Kondo coupling $J_{cf}$ becomes weaker.
Our findings provide microscopic evidence that chemical substitution in Ce(Pt$_{1-x}$Pd$_{x}$)$_6$Al$_3$ enables controlled tuning between itinerant and localized magnetism.

\subsection{Prospects for Unconventional Superconductivity}
Although superconductivity has not yet been observed in Ce(Pt$_{1-x}$Pd$_{x}$)$_6$Al$_3$, the proximity to a magnetic QCP near $x_{\mathrm{c}} \simeq 0$ raises the intriguing possibility of its emergence.
Unconventional superconductivity often appears in the vicinity of a magnetic QCP, where magnetic fluctuations can mediate unconventional pairing mechanisms~\cite{C.Pfleiderer_RMP_2009,Y.Nakai_PRL_2010,S.Kitagawa_PRB_2019,K.Ishida_PRB_2021}.
Given the evidence for a QCP and the suppression of the Kondo scale with Pd substitution, Ce(Pt$_{1-x}$Pd$_{x}$)$_6$Al$_3$ represents a promising candidate for the realization of such superconductivity.
Furthermore, the presence of competing interactions in the honeycomb network of Ce ions could enrich the pairing landscape, potentially leading to novel superconducting states that intertwine with magnetic frustration~\cite{PhysRevB.86.100507,Watanabe_JPSJ_2017,tazaiMechanismExoticDensitywave2022}.
To explore this possibility, future studies with finer control of the Pd concentration near $x_{\mathrm{c}}$, higher sample purity, and application of external pressure would be highly desirable.
Moreover, the combination of frustration and quantum criticality in Ce(Pt$_{1-x}$Pd$_{x}$)$_6$Al$_3$ offers a rare opportunity to investigate the largely unexplored relationship between magnetic frustration and unconventional superconductivity in correlated electron systems.

\section{Conclusion}
We have performed comprehensive $^{27}$Al and $^{195}$Pt nuclear magnetic resonance measurements on polycrystalline Ce(Pt$_{1-x}$Pd$_{x}$)$_6$Al$_3$ with $x = 0$, 0.1, 0.2, and 0.3 to investigate the evolution of magnetic properties as a function of Pd substitution.
The Knight shift, linewidth, and nuclear spin--lattice relaxation rate collectively reveal that CePt$_6$Al$_3$ remains a paramagnetic heavy-fermion metal down to 0.1~K, characterized by $T_{\mathrm{coh}} \simeq 15$~K.
In contrast, Pd substitution at $x \geq 0.1$ induces long-range AFM order below $T_{\mathrm{N}} \simeq 3.5$~K, accompanied by a suppression of $T_{\mathrm{coh}}$.
The distinct behaviors of $T_{\mathrm{coh}}$ and $T_{\mathrm{N}}$ imply that Pd doping not only reduces the Kondo coupling $J_{cf}$ but also partially relieves the underlying magnetic frustration.
A QCP is suggested to exist near $x_{\mathrm{c}} \simeq 0$.
Comparison between $x = 0.1$ and $x = 0.3$ reveals a crossover from itinerant SDW–type to localized moment antiferromagnetism as the Kondo effect weakens.
These findings establish Ce(Pt$_{1-x}$Pd$_{x}$)$_6$Al$_3$ as a unique platform where the interplay between Kondo screening and magnetic frustration can be systematically tuned by chemical substitution.
The combination of magnetic frustration and quantum criticality provides an exciting opportunity to explore the potential emergence of unconventional superconductivity and exotic quantum phases in this system.

\section*{acknowledgments}
This work was supported by Grants-in-Aid for Scientific Research (KAKENHI Grant No. JP20KK0061, No. JP20H00130, No. JP21K03473, No. JP21K18600, No. JP22H04933, No. JP22H01168, No. JP22KJ2336, No. JP23H01124, No. JP23K22439 and No. JP23K25821) from the Japan Society for the Promotion of Science, by JST SPRING(Grant No. JPMJSP2110) from the Japan Science and Technology Agency, by research support funding from the Kyoto University Foundation, by ISHIZUE 2024 of Kyoto University Research Development Program, by Murata Science and Education Foundation, and by the JGC-S Scholarship Foundation.
Liquid helium is supplied by the Low Temperature and Materials Sciences Division, Agency for Health, Safety and Environment, Kyoto University.

\end{document}